\documentclass[12pt,a4paper,twoside]{article}
\usepackage{amsmath}
\usepackage{amsfonts}
\usepackage{amssymb}
\usepackage{amsfonts}
\newcommand{\widebar}[1]{\mkern 1.5mu\overline{\mkern-1.5mu#1\mkern-1.5mu}\mkern 1.5mu}
\usepackage{mathtools}
\newenvironment{changemargin}[2]{%
\begin{list}{}{%
\setlength{\leftmargin}{#1}%
\setlength{\rightmargin}{#2}%
}%
\item[]}
{\end{list}}
\usepackage{graphicx}
\usepackage{color}
\usepackage{lscape}
\usepackage{epstopdf}
\usepackage{lscape}
\usepackage{amsmath}
\usepackage{amsmath}
\usepackage{setspace}

\begin{document}
\baselineskip=0.25in
{\bf \LARGE
\begin{changemargin}{-0.5cm}{-0.5cm}
\begin{center}
{Teleportation with two-dimensional electron gas formed at the interface of a GaAs heterostructure}\footnote{\scriptsize {\bf To appear in Laser Phys. 2017}}
\end{center}\end{changemargin}}
\vspace{4mm}
\begin{center}
\large{\bf Adenike Grace Adepoju$^{a,}$}\footnote{\scriptsize E-mail:~ aagrace06@gmail.com}\large{\bf ,} \large{\bf Babatunde James Falaye$^{b,c,}$}\footnote{\scriptsize E-mail:~ fbjames11@physicist.net; babatunde.falaye@fulafia.edu.ng}\large{\bf ,} {\large{\bf Guo-Hua Sun$^{d,}$}}\footnote{\scriptsize E-mail:~ sunghdb@yahoo.com}\large{\bf ,} {\large{\bf Oscar Camacho-Nieto$^{a,}$}}\footnote{\scriptsize E-mail:~ ocamacho@ipn.mx} \large{\bf and} {\large{\bf Shi-Hai Dong$^{a,}$}}\footnote{\scriptsize E-mail:~ dongsh2@yahoo.com }
\end{center}
{\footnotesize
\begin{center}
{\it $^\textbf{a}$CIDETEC, Instituto Polit\'{e}cnico Nacional, UPALM, CDMX 07700, M\'{e}xico.}\\
{\it $^\textbf{b}$Departamento de F\'isica, Escuela Superior de F\'isica y Matem\'aticas, Instituto Polit\'ecnico Nacional, Edificio 9, UPALM, CDMX 07738, M\'{e}xico.}\\
{\it $^\textbf{c}$Applied Theoretical Physics Division, Department of Physics, Federal University Lafia,  P. M. B. 146, Lafia, Nigeria.}\\
{\it $^\textbf{d}$C\'{a}tedr\'{a}tica CONACyT, CIC, Instituto Polit\'{e}cnico Nacional, UPALM, CDMX 07700, M\'{e}xico.}
\end{center}}
\begin{abstract}
\noindent
Inspired by the scenario by Bennett et al., a teleportation protocol of qubits formed in a two-dimensional electron gas formed at the interface of a GaAs heterostructure is presented. The teleportation is carried out using three GaAs quantum dots (say $\mathcal{P}\mathcal{P}'$, $\mathcal{Q}\mathcal{Q}'$, $\mathcal{R}\mathcal{R}'$) and three electrons. The electron spin on GaAs quantum dots $\mathcal{P}\mathcal{P}'$ is used to encode the unknown qubit. The GaAs quantum dot $\mathcal{Q}\mathcal{Q}'$ and $\mathcal{R}\mathcal{R}'$ combine to form {an} entangled state. Alice (the sender) performs a Bell measurement on pairs ($\mathcal{P},\mathcal{Q}$) and ($\mathcal{P}',\mathcal{Q}'$). Depending on the outcome of {the} measurement, a suitable Hamiltonian for the quantum gate can be used by Bob (receiver) to transform the information based on {a} spin to charge-based information. This work offers relevant corrections to misconception in {\it Chem. Phys. Lett. {\bf421} (2006) 338.}
\noindent
\end{abstract}


{\bf PACs No.}: 03.65.Ud, 03.67.Hk, 03.67.Ac, 03.67.Mn.
\vspace{4mm}

\section{Introduction}
In quantum teleportation, the whole object positioned at a point, say $A$ cannot be teleported, but only its state with the aid of classical communication and previously shared entanglement between points $A$ and another point, say $B$. This idea was expounded in paper of  Bennett et al. \cite{A1}. A description of the standard teleportation protocol is as follows: The sender Alice has a source qubit, say $\left|\Phi\right\rangle_{A'}=a\left|0\right\rangle+b\left|1\right\rangle$ which she wants to teleport to Bob. Suppose Alice and Bob share another qubit. Then, the entangled state of the two parties could be represented by $\left|\Psi \right\rangle_{AB}={1}/{\sqrt{2}}\left(|\left.0\right\rangle_A\otimes|\left.0\right\rangle_B+|\left.1\right\rangle_A\otimes|\left.1\right\rangle_B\right)$ , where the $_A$ and $_B$ {denote} the quantum state of Alice and Bob respectively. Since Alice does not know the state of $\left|\Phi\right\rangle_{A'}$ (the laws of quantum mechanics do not permit this since she only has a single copy of $\left.|\Psi\right\rangle_{A'}$ in her possession), therefore, it is impossible for her to precisely measure and specify it. Now, Alice relates the qubit in her possession with half of the EPR pair (i.e., $\left.|\Phi^{(0)}\right\rangle_{A'AB}=\left|\Phi\right\rangle_{A'}\left|\Psi\right\rangle_{AB}$, where $\left|\Phi^{(0)}\right\rangle_{A'AB}$ denotes the state input to the circuit) and sends the qubits through a C-NOT gate to obtain $\left|\Phi^{(1)}\right\rangle_{A'AB}=\frac{1}{\sqrt{2}}\left[a\left.|0\right\rangle\left(\left.|00\right\rangle+\left.|11\right\rangle\right)+b\left.|1\right\rangle\left(\left.|10\right\rangle+\left.|01\right\rangle\right)\right]$ and then sends the first qubit through a Hadamard gate. The result of this becomes
\begin{equation} \left|\Phi^{(2)}\right\rangle_{A'AB}=\frac{1}{{2}}\left[\left|00\right\rangle\left(\left.a|0\right\rangle+\left.b|1\right\rangle\right)+\left|01\right\rangle\left(\left.a|1\right\rangle+\left.b|0\right\rangle\right)+\left|10\right\rangle\left(\left.a|0\right\rangle-\left.b|1\right\rangle\right)+\left|11\right\rangle\left(\left.a|1\right\rangle-\left.b|0\right\rangle\right)\right].
\label{E1}
\end{equation}
From the equation (\ref{E1}), one can deduce that each state of Alice's qubits corresponds to a state of Bob's qubit. Once Bob gets the result of this measurement, he can reconstructs the state $\left|\Phi\right\rangle_{A'}$ by applying the appropriate quantum gate. As it has just been demonstrated, a quantum state can be recovered in a remote location with the aid of a maximally entangled  EPR pair and two bits of classical information. On the basis of this novel idea, there has been a continuous interest in studying quantum teleportation. The literature is huge, however, the following  references \cite{AS2, AS4, AS5, QW1, AS6, KL1, KL2} give some older as well as recent studies.
	
Moreover, the usefulness of entanglement is not restricted to teleportation alone, it also forms an imperative component in quantum cryptography and quantum information sharing.  All these justify indispensability of an entanglement in quantum information processing. It is a manifestation of intrinsic non-locality in {the} quantum mechanic. A comprehensive understanding of {behavior} of entangled systems in their environs has been an investigation which is common to quantum measurement and quantum information. To the best of our knowledge, no much work on teleportation in semiconductor has so far been achieved. Nanoscale systems such as quantum dots and superconducting circuits make good candidates of standard semiconductor technology for practical quantum computers. Within this {context}, a quantum teleportation  might be a decisive confirmation of its potentiality.

A quantum dot is a semiconductor nanostructure that incarcerates (confinement can be as a result of the presence of electrostatic potentials, semiconductor surface, an interface between different semiconductor materials) the motion of conduction band electrons, valence band holes, or excitons (bound pairs of conduction band electrons and valence band holes) in all three spatial directions. A crucial way forward in quantum computation was  the realization of {a} dot in GaAs \cite{AS7}. In GaAs quantum dots, electron spins are used as qubits. The qubits are formed in a standard two-dimensional electron gas (2DEG) obtained at the interface of a GaAs/AlGaAs heterostructure. For  realization of well-defined spin qubits, the model equation for the on-site ($U$) and nearest-neighbor ($U_{12}$) Coulomb repulsion, respectively in a magnetic field is given by the Hamiltonian
\begin{equation}
H=H_0+V_m=\frac{1}{2}\sum_{i}UN_i(N_i-1)+U_{12}N_1N_2-e\sum{V_iN_i}+\sum_{i,k}\epsilon_{ik}n_{ik}-\vec{\mu}\cdot\vec{B},
\label{E2}
\end{equation}
where $H_0$ is the unperturbed Hamiltonian \cite{AS8} and $V_m$ is correction to $H_0$ which results from magnetic fields and spin-couplings.  $\vec{\mu}=g\mu_B\sum_{m}S_m$ is the magnetic moment with $\mu_B$ being the Bohr magneton. The $m$-th electron's spin-1/2 in the double dot is denoted by $S_m$.  $g$ is the Landa\'{e} g-factor and $N_i=\sum_{k}n_{ik}$ counts the total number of electrons in dot $i$ with $n_{ik}=\sum_{\sigma}d_{ik\sigma}^\dag d_{ik\sigma}$. $d_{ik\sigma}$ annihilates an electron on dot $i$, in orbital $k$ with spin $\sigma$.  $\epsilon_{ik}$ denotes the energy of single-particle orbital level in dot, which yields the typical orbital level spacing $\epsilon_{ik+1}-\epsilon_{ik}\approx\hbar\omega_0$. 

With a sufficiently large magnetic field intensity, {a} system of qubit can be initialized. Xu et al. \cite{A2} initialized a spin state with a singly charged InAs-GaAs quantum dot by optical cooling with near unity efficiency. The experiment requires magnetic field of 0.88 T in Voigt geometry and temperature of 5-0.06K. The details can be read in ref. \cite{A2}. Very recently Mar et al. \cite{A3} demonstrated that without the need of magnetic field, the initialization of a single quantum-dot hole spin with high fidelity (lower bound $>97\%$), on picosecond time scales could as well be {realized}. The spin configuration has relevant eigenstates corresponding to $|\left.\uparrow\downarrow\right\rangle$ or $|\left.\downarrow\uparrow\right\rangle$

By occupying each dot with exactly one electron, the spin qubits are realized. Generally speaking, there are many systems that could be employed as qubits in a quantum computation such as the polarization of a single photon (the two states are vertical polarization and horizontal polarization). In this work, we demonstrate a quantum teleportation protocol by utilizing the spins of electrons confined in GaAs quantum dots. The teleportation is carried out using three GaAs quantum dots (say $\mathcal{P}\mathcal{P}'$, $\mathcal{Q}\mathcal{Q}'$, $\mathcal{R}\mathcal{R}'$) and three electrons. The electron spin on GaAs quantum dot $\mathcal{P}\mathcal{P}'$ is used to encode the unknown qubit. The GaAs quantum dots $\mathcal{Q}\mathcal{Q}'$ and $\mathcal{R}\mathcal{R}'$ combine to form {an} entangled state. We perform a Bell measurement and depending on the outcome of the measurement, a suitable Hamiltonian for quantum gate can be used to transform the spin-base information to charge-base information.

\section{Demonstration of quantum teleportation in quantum dot}
\begin{figure*}[!h]
\centering \includegraphics[height=55mm, width=130mm]{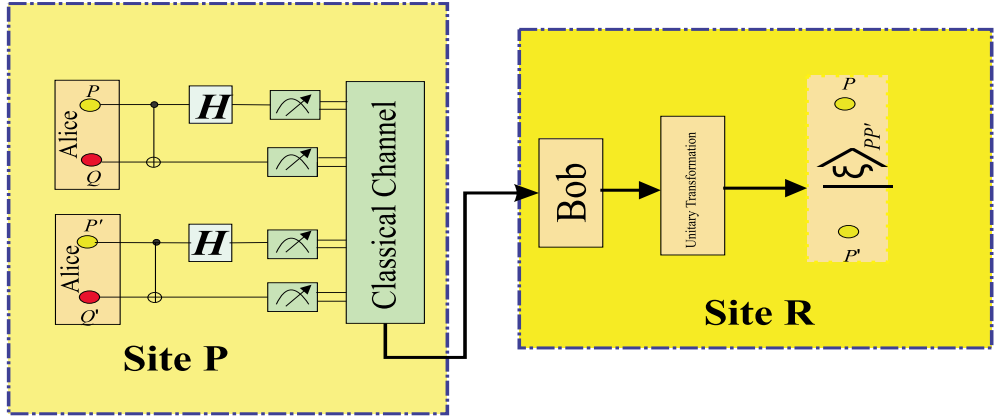}
\caption{\protect\footnotesize Schematic representation of quantum teleportation protocol with two-dimensional electron gas formed at the interface of a GaAs heterostructure. The meters M$_1$ and M$_2$ represent the measurement. The particles $a$, $b$, $1$ and $6$ belongs to Alice, while particles $2$ and $3$ belongs to Bob and Particles $4$ and $5$ to Chika. $H$ is the Hadamard gate. The double lines coming out of the meters carry classical bits (single lines denote qubits). Alice performs Bell state measurement on her qubit pairs $(\mathcal{P},\mathcal{Q})$ and $(\mathcal{P}',\mathcal{Q}')$. Alice communicates the results of her measurement to Bob via classical channel. With information received from from Alice, Bob can recover the original state of the qubit (also in the same color combo) via an appropriate unitary transformation at site $R$.}
\label{fig1}
\end{figure*}

Now, let there be two participants, spatially separated in different sites in a quantum network, customarily called Alice and Bob. The qubit that Alice, who is located at site $P$, wishes to teleport to Bob at site ${R}$, has been obtained in a standard 2DEG formed at the interface of a GaAs/AlGaAs heterostructure, and it can be written as $\left|\xi\right\rangle_{\mathcal{P}\mathcal{P}'}=\left.\alpha|\downarrow\uparrow\right\rangle+\left.\beta|\uparrow\downarrow\right\rangle$, where $\alpha$ and $\beta$ are complex and satisfy the relation $|\alpha|^2 + |\beta|^2 = 1$. The entangled state of $\mathcal{Q}$ and $\mathcal{R}$ can be written in occupation number basis $\left|n_{\mathcal{Q}}\uparrow n_{\mathcal{Q}'}\downarrow n_{\mathcal{R}}\uparrow n_{\mathcal{R}'}\downarrow\right\rangle$
\begin{equation}
\left|\chi\right\rangle_{\mathcal{Q}\mathcal{Q}'\mathcal{R}\mathcal{R}'}=\frac{1}{2}\left(\left|0001\right\rangle+\left|0100\right\rangle+\left|1011\right\rangle+\left|1110\right\rangle\right).
\label{E3}
\end{equation}
Qubits pair $(\mathcal{Q}\mathcal{Q}')$ belong to Alice while $(\mathcal{R}\mathcal{R}')$ belong to Bob. It can be deduced from equation (\ref{E3}) that, there are four possible states in $\mathcal{Q}\mathcal{Q}'$ which correspond to each state in $\mathcal{R}\mathcal{R}'$, i.e. ($\left|00\right\rangle\Leftrightarrow\left|01\right\rangle$, $\left|01\right\rangle\Leftrightarrow\left|00\right\rangle$, $\left|10\right\rangle\Leftrightarrow\left|11\right\rangle$ and $\left|11\right\rangle\Leftrightarrow\left|10\right\rangle$). The combined state of the qubits becomes $\left|\Phi^{(0)}\right\rangle_{\mathcal{P}\mathcal{P}'\mathcal{Q}\mathcal{Q}'\mathcal{R}\mathcal{R}'}=\left|\xi\right\rangle_{\mathcal{P}\mathcal{P}'}\left|\chi\right\rangle_{\mathcal{Q}\mathcal{Q}'\mathcal{R}\mathcal{R}'}$. For Alice to achieve her aim, she performs a Bell state measurement on her qubits pair $(\mathcal{P}\mathcal{Q})$ to obtain the states of other qubits as
\begin{subequations}
\begin{eqnarray}
_{\mathcal{P}\mathcal{Q}}\left\langle\Phi^{\pm}\right.\left|\Phi^{(0)}\right\rangle_{\mathcal{P}\mathcal{P}'\mathcal{Q}\mathcal{Q}'\mathcal{R}\mathcal{R}'}=\frac{1}{2\sqrt{2}}\left[\pm^{(1)}\alpha\left(\left|0011\right\rangle+\left|0110\right\rangle\right)+ \beta\left(\left|1001\right\rangle+\left|1100\right\rangle\right)\right]=\left|\Phi^{(1a)}\right\rangle_{\mathcal{P}'\mathcal{Q}'\mathcal{R}\mathcal{R}'},\label{EQ4a}\\
_{\mathcal{P}\mathcal{Q}}\left\langle\Psi^{\pm}\right.\left|\Phi^{(0)}\right\rangle_{\mathcal{P}\mathcal{P}'\mathcal{Q}\mathcal{Q}'\mathcal{R}\mathcal{R}'}=\frac{1}{2\sqrt{2}}\left[\pm^{(1)}\alpha\left(\left|0001\right\rangle+\left|0100\right\rangle\right)+ \beta\left(\left|1011\right\rangle+\left|1110\right\rangle\right)\right]=\left|\Phi^{(1b)}\right\rangle_{\mathcal{P}'\mathcal{Q}'\mathcal{R}\mathcal{R}'}\label{EQ4b},
\end{eqnarray}
\end{subequations}
and then on qubit pair $(\mathcal{P}'\mathcal{Q}')$. Thus, the states of qubits pair $(\mathcal{R}\mathcal{R}')$ become
\begin{subequations}
\begin{eqnarray}
&& _{\mathcal{P}'\mathcal{Q}'}\left\langle\Phi^{\pm}\right|\ _{\mathcal{P}\mathcal{Q}}\left\langle\Phi^{\pm}\right|\Phi^{(0)}\big\rangle_{\mathcal{P}\mathcal{P}'\mathcal{Q}\mathcal{Q}'\mathcal{R}\mathcal{R}'}=\frac{1}{4}\Big[\pm^{(1)}\alpha\left|\downarrow\downarrow\right\rangle\pm^{(2)}\beta\left|\uparrow\uparrow\right\rangle\Big]=\left|\Phi^{(2a)}\right\rangle_{\mathcal{R}\mathcal{R}'}\label{EQ5a}, \\
&& _{\mathcal{P}'\mathcal{Q}'}\left\langle\Psi^{\pm}\right|\ _{\mathcal{P}\mathcal{Q}}\left\langle\Phi^{\pm}\right|\Phi^{(0)}\big\rangle_{\mathcal{P}\mathcal{P}'\mathcal{Q}\mathcal{Q}'\mathcal{R}\mathcal{R}'}=\frac{1}{4}\Big[\pm^{(1)}\alpha\left|\downarrow\uparrow\right\rangle\pm^{(2)}\beta\left|\uparrow\downarrow\right\rangle\Big]=\left|\Phi^{(2b)}\right\rangle_{\mathcal{R}\mathcal{R}'}\label{EQ5b},\\
&& _{\mathcal{P}'\mathcal{Q}'}\left\langle\Phi^{\pm}\right|\ _{\mathcal{P}\mathcal{Q}}\left\langle\Psi^{\pm}\right|\Phi^{(0)}\big\rangle_{\mathcal{P}\mathcal{P}'\mathcal{Q}\mathcal{Q}'\mathcal{R}\mathcal{R}'}=\frac{1}{4}\Big[\pm^{(1)}\alpha\left|\uparrow\downarrow\right\rangle\pm^{(2)}\beta\left|\downarrow\uparrow\right\rangle\Big]=\left|\Phi^{(2c)}\right\rangle_{\mathcal{R}\mathcal{R}'}\label{EQ5c}, \\
&& _{\mathcal{P}'\mathcal{Q}'}\left\langle\Psi^{\pm}\right|\ _{\mathcal{P}\mathcal{Q}}\left\langle\Psi^{\pm}\right|\Phi^{(0)}\rangle_{\mathcal{P}\mathcal{P}'\mathcal{Q}\mathcal{Q}'\mathcal{R}\mathcal{R}'}=\frac{1}{4}\Big[\pm^{(1)}\alpha\left|\uparrow\uparrow\right\rangle\pm^{(2)}\beta\left|\downarrow\downarrow\right\rangle\Big]=\left|\Phi^{(2d)}\right\rangle_{\mathcal{R}\mathcal{R}'}\label{EQ5d},
\end{eqnarray}
\end{subequations}
\begin{table}[!h]
\caption{\footnotesize Alice's results, the corresponding state obtained by Bob and the appropriate unitary transformation which can be utilized by Bob to reconstruct the original state of the qubit. We have ignored the normalization for convenience} \vspace*{10pt}{\footnotesize
\begin{tabular}{ccccccc}\hline\hline
{}&{}&{}&{}&{}&{}&{}\\[-1.0ex]Alice's result&&& State obtained by Bob&&& Unitary transformation\\[1ex]\hline
$\left|\Phi^+\right\rangle\left|\Phi^+\right\rangle$&&&$\alpha\left|\downarrow\downarrow\right\rangle+\beta\left|\uparrow\uparrow\right\rangle$&&&$\left(\left|\uparrow\right\rangle\left\langle\uparrow\right|+\left|\downarrow\right\rangle\left\langle\downarrow\right|\right)\otimes \left(\left|\uparrow\right\rangle\left\langle\downarrow\right|+\left|\downarrow\right\rangle\left\langle\uparrow\right|\right)$ \\[1ex]
$\left|\Phi^+\right\rangle\left|\Phi^-\right\rangle$&&&$-\alpha\left|\downarrow\downarrow\right\rangle+\beta\left|\uparrow\uparrow\right\rangle$&&&$\left(\left|\uparrow\right\rangle\left\langle\uparrow\right|-\left|\downarrow\right\rangle\left\langle\downarrow\right|\right)\otimes \left(\left|\uparrow\right\rangle\left\langle\downarrow\right|+\left|\downarrow\right\rangle\left\langle\uparrow\right|\right)$ \\[1ex]
$\left|\Phi^-\right\rangle\left|\Phi^+\right\rangle$&&&$\alpha\left|\downarrow\downarrow\right\rangle-\beta\left|\uparrow\uparrow\right\rangle$&&&$\left(\left|\uparrow\right\rangle\left\langle\uparrow\right|+\left|\downarrow\right\rangle\left\langle\downarrow\right|\right)\otimes\left(\left|\uparrow\right\rangle\left\langle\downarrow\right|-\left|\downarrow\right\rangle\left\langle\uparrow\right|\right)$ \\[1ex]
$\left|\Phi^-\right\rangle\left|\Phi^-\right\rangle$&&&$-\alpha\left|\downarrow\downarrow\right\rangle-\beta\left|\uparrow\uparrow\right\rangle$&&&$\left(\left|\uparrow\right\rangle\left\langle\uparrow\right|-\left|\downarrow\right\rangle\left\langle\downarrow\right|\right)\otimes\left(\left|\uparrow\right\rangle\left\langle\downarrow\right|-\left|\downarrow\right\rangle\left\langle\uparrow\right|\right)$ \\[0.5ex]\hline
$\left|\Psi^+\right\rangle\left|\Phi^+\right\rangle$&&&$\alpha\left|\downarrow\uparrow\right\rangle+\beta\left|\uparrow\downarrow\right\rangle$&&&$\left(\left|\uparrow\right\rangle\left\langle\uparrow\right|+\left|\downarrow\right\rangle\left\langle\downarrow\right|\right)\otimes\left(\left|\uparrow\right\rangle\left\langle\uparrow\right|+\left|\downarrow\right\rangle\left\langle\downarrow\right|\right)$ \\[1ex]
$\left|\Psi^+\right\rangle\left|\Phi^-\right\rangle$&&&$-\alpha\left|\downarrow\uparrow\right\rangle+\beta\left|\uparrow\downarrow\right\rangle$&&&$\left(\left|\uparrow\right\rangle\left\langle\uparrow\right|-\left|\downarrow\right\rangle\left\langle\downarrow\right|\right)\otimes \left(\left|\uparrow\right\rangle\left\langle\uparrow\right|+\left|\downarrow\right\rangle\left\langle\downarrow\right|\right)$ \\[1ex]
$\left|\Psi^-\right\rangle\left|\Phi^+\right\rangle$&&&$\alpha\left|\downarrow\uparrow\right\rangle-\beta\left|\uparrow\downarrow\right\rangle$&&&$\left(\left|\uparrow\right\rangle\left\langle\uparrow\right|+\left|\downarrow\right\rangle\left\langle\downarrow\right|\right)\otimes \left(\left|\uparrow\right\rangle\left\langle\uparrow\right|-\left|\downarrow\right\rangle\left\langle\downarrow\right|\right)$ \\[1ex]
$\left|\Psi^-\right\rangle\left|\Phi^-\right\rangle$&&&$-\alpha\left|\downarrow\uparrow\right\rangle-\beta\left|\uparrow\downarrow\right\rangle$&&&$\left(\left|\uparrow\right\rangle\left\langle\uparrow\right|-\left|\downarrow\right\rangle\left\langle\downarrow\right|\right)\otimes\left(\left|\uparrow\right\rangle\left\langle\uparrow\right|-\left|\downarrow\right\rangle\left\langle\downarrow\right|\right)$  \\[0.5ex]\hline
$\left|\Phi^+\right\rangle\left|\Psi^+\right\rangle$&&&$\alpha\left|\uparrow\downarrow\right\rangle+\beta\left|\downarrow\uparrow\right\rangle$&&&$\left(\left|\uparrow\right\rangle\left\langle\downarrow\right|-\left|\downarrow\right\rangle\left\langle\uparrow\right|\right)\otimes\left(\left|\downarrow\right\rangle\left\langle\uparrow\right|-\left|\uparrow\right\rangle\left\langle\downarrow\right|\right)$ \\[1ex]
$\left|\Phi^+\right\rangle\left|\Psi^-\right\rangle$&&&$-\alpha\left|\uparrow\downarrow\right\rangle+\beta\left|\downarrow\uparrow\right\rangle$&&&$\left(\left|\uparrow\right\rangle\left\langle\downarrow\right|-\left|\downarrow\right\rangle\left\langle\uparrow\right|\right)\otimes \left(\left|\uparrow\right\rangle\left\langle\downarrow\right|+\left|\downarrow\right\rangle\left\langle\uparrow\right|\right)$ \\[1ex]
$\left|\Phi^-\right\rangle\left|\Psi^+\right\rangle$&&&$\alpha\left|\uparrow\downarrow\right\rangle-\beta\left|\downarrow\uparrow\right\rangle$&&&$\left(\left|\downarrow\right\rangle\left\langle\uparrow\right|-\left|\uparrow\right\rangle\left\langle\downarrow\right|\right)\otimes \left(\left|\uparrow\right\rangle\left\langle\downarrow\right|+\left|\downarrow\right\rangle\left\langle\uparrow\right|\right)$ \\[1ex]
$\left|\Phi^-\right\rangle\left|\Psi^-\right\rangle$&&&$-\alpha\left|\uparrow\downarrow\right\rangle-\beta\left|\downarrow\uparrow\right\rangle$&&&$\left(\left|\uparrow\right\rangle\left\langle\downarrow\right|-\left|\downarrow\right\rangle\left\langle\uparrow\right|\right)\otimes\left(\left|\downarrow\right\rangle\left\langle\uparrow\right|-\left|\uparrow\right\rangle\left\langle\downarrow\right|\right)$ \\[0.5ex]\hline
$\left|\Psi^+\right\rangle\left|\Psi^+\right\rangle$&&&$\alpha\left|\uparrow\uparrow\right\rangle+\beta\left|\downarrow\downarrow\right\rangle$&&&$\left(\left|\downarrow\right\rangle\left\langle\uparrow\right|-\left|\uparrow\right\rangle\left\langle\downarrow\right|\right)\otimes \left(\left|\uparrow\right\rangle\left\langle\uparrow\right|-\left|\downarrow\right\rangle\left\langle\downarrow\right|\right)$ \\[1ex]
$\left|\Psi^+\right\rangle\left|\Psi^-\right\rangle$&&&$-\alpha\left|\uparrow\uparrow\right\rangle+\beta\left|\downarrow\downarrow\right\rangle$&&&$\left(\left|\uparrow\right\rangle\left\langle\downarrow\right|-\left|\downarrow\right\rangle\left\langle\uparrow\right|\right)\otimes \left(\left|\uparrow\right\rangle\left\langle\uparrow\right|+\left|\downarrow\right\rangle\left\langle\downarrow\right|\right)$ \\[1ex]
$\left|\Psi^-\right\rangle\left|\Psi^+\right\rangle$&&&$\alpha\left|\uparrow\uparrow\right\rangle-\beta\left|\downarrow\downarrow\right\rangle$&&&$\left(\left|\downarrow\right\rangle\left\langle\uparrow\right|-\left|\uparrow\right\rangle\left\langle\downarrow\right|\right)\otimes \left(\left|\uparrow\right\rangle\left\langle\uparrow\right|+\left|\downarrow\right\rangle\left\langle\downarrow\right|\right)$ \\[1ex]
$\left|\Psi^-\right\rangle\left|\Psi^-\right\rangle$&&&$-\alpha\left|\uparrow\uparrow\right\rangle-\beta\left|\downarrow\downarrow\right\rangle$&&&$\left(\left|\uparrow\right\rangle\left\langle\downarrow\right|-\left|\downarrow\right\rangle\left\langle\uparrow\right|\right)\otimes \left(\left|\uparrow\right\rangle\left\langle\uparrow\right|-\left|\downarrow\right\rangle\left\langle\downarrow\right|\right)$ \\[0.5ex]\hline\hline
\end{tabular}\label{tab1}}
\end{table}
where we have denoted the four Bell states as $\left|\Phi^{\pm}\right\rangle=2^{-1/2}(\left|\uparrow\uparrow\right\rangle\pm\left|\downarrow\downarrow\right\rangle)$ and $\left|\Psi^{\pm}\right\rangle=2^{-1/2}(\left|\uparrow\downarrow\right\rangle\pm\left|\downarrow\uparrow\right\rangle)$. The $\pm^{(1)}$ and $\pm^{(2)}$  represent the results corresponding to BSM on qubit pair ($\mathcal{P},\mathcal{Q}$) and ($\mathcal{P}',\mathcal{Q}'$) respectively. The above states represent the $16$ possible states which Alice's system will collapse into after the measurement. One can also observe that these states are pure entangled two-qubit states. Alice then communicates the results of the measurement to Bob, who can choose an apt unitary transformation via the electron spin up and spin down as basis in order to completely recovers $\left|\xi\right\rangle_{\mathcal{P}\mathcal{P}'}$ on site $R$. Thus, the {spin-based} information has undergo a transformation to charge based information via a proper Hamiltonian for the quantum gate. Moreover, the content of the information is unaffected. A  more explicit expression for equations (\ref{EQ5a}-\ref{EQ5d}) and the appropriate unitary transformation which Bob could utilize in reconstructing the original state, are shown in Table 1. This process is also illustrated in Figure \ref{fig1}.

It is worth mentioning that because of {the} nonlinearity of interactions which is involved in the model, the only {obstacle} which could incapacitate our Bell {measurement} from reaching $100\%$ of {success} probability is noise. Noise may set in while Alice performs the Bell measurement and Bob  perform the unitary operation. This might be due to {an} imperfection of local operation. The properties of teleportation through noisy quantum channels can be measured by the fidelity. The most significant model for noise is depolarizing channel, which is known to introduce white noise. The state of quantum system after transmitted through a depolarizing channel can be written as $\mathcal{E}_d(\hat{\rho}) = pI/d+ (1-p)\hat{\rho}$, where $\hat{\rho}$ is a general mixed state representing the initial state of the qubit and $d$ denotes the dimension of Hilbert space. This expression can be described by saying that, with probability $p$, an error occurs while  with probability $1-p$, the qubit remains intact.

\section{Conclusion}
In this paper, we have presented a model of quantum teleportation protocol by utilizing the spins of electrons confined in GaAs quantum dots to perform a teleportation. We have utilized three systems of electrons ($\mathrm{e^-}=3$). However, for $\mathrm{e^-}>3$, where on-site Coulomb repulsion approaches positive infinity, there will be no double occupation and the antiparallel configuration will be favored by the neighboring spin. Moreover, the current work has corrected misconception in Ref. \cite{AS2}. The authors performed a wrong Bell state measurement. We will not go into details of this anyway. However, We would like to point the attention of the readers to Eq. (16) of ref. \cite{AS2} which is the output of Hadamard transformation. The Hadamard transform for 2 qubits is widely known to be
\begin{equation}
H_2=\frac{1}{2}\begin{pmatrix*}[r]1&1&1&1\\1&-1&1&-1&\\1&1&-1&-1\\1&-1&-1&1\end{pmatrix*}.
\end{equation}
Obviously, its operation onto $\alpha\left|10\right\rangle$ should be
\begin{equation}
\frac{1}{2}\begin{pmatrix*}[r]1&1&1&1\\1&-1&1&-1&\\1&1&-1&-1\\1&-1&-1&1\end{pmatrix*}\begin{pmatrix*}[r]0\\0\\ \alpha\\0\end{pmatrix*}=\frac{1}{2}\begin{pmatrix*}[r]\alpha\\ \alpha\\ -\alpha\\-\alpha\end{pmatrix*}=\frac{\alpha}{2}\left(\left|00\right\rangle+\left|01\right\rangle-\left|10\right\rangle-\left|11\right\rangle\right),
\end{equation}
while for $\beta\left|01\right\rangle$ should be
${\beta}/{2}\left(\left|00\right\rangle-\left|01\right\rangle+\left|10\right\rangle-\left|11\right\rangle\right)$. Supposing that the idea of Alice performing BSM on qubits pair $(AACC)$ is right, then, one can clearly see that Eq. (16) in Ref. \cite{AS2} is wrong and the correct expression should have been
\begin{equation}
\left|\Psi_2\right\rangle=\frac{\alpha}{2\sqrt{2}}\left(\left|00\right\rangle+\left|01\right\rangle-\left|10\right\rangle-\left|11\right\rangle\right)(\left|0000\right\rangle+\left|1111\right\rangle)\nonumber+\frac{\beta}{2\sqrt{2}}\left(\left|00\right\rangle-\left|01\right\rangle+\left|10\right\rangle-\left|11\right\rangle\right)(\left|1100\right\rangle+\left|0011\right\rangle).
\end{equation}
The authors proceeded with the wrong expression (16) to obtain (17) which is another flaw idea. The well known circuit for BSM can be found in page 26 of Ref. \cite{BJ2}. Now, this implies that if we are to follow their idea by performing BSM on qubits pair $(AACC)$, then $x$ would be $AA$ and y would be $CC$ which is not meaningful (see equation 1.27 of Ref. \cite{BJ2} for meaning of $x$ and $y$). For instance, let us consider the pair $\left|1011\right\rangle$ in Ref. \cite{AS2}. Then, using Eq. 1.27 of Ref. \cite{BJ2}, one will obtain $\left|\beta_{10,11}\right\rangle=(\left|0,11\right\rangle+(-1)^{10}\left|1,\widebar{11}\right\rangle)/\sqrt{2}$ which is not meaningful.  The same approach was also applied to ebit $\beta_1$ as shown in Eqs. (19), (20), (21) and (22) of Ref. \cite{AS2}. The approach of the current study has corrected this misconception.

\section*{Acknowledgments}
We thank the referees for the positive enlightening comments and suggestions, which have greatly helped us in making improvements to this paper. This work was partially supported by 20160978-SIP-IPN, Mexico.


\begin{thebibliography}{99} 
\bibitem{A1} C. H. Bennett, G. Brassard, C. Crepeau, R. Jozsa, A. Peres, W. K. Wootters, "Teleporting an Unknown Quantum State via Dual Classical and Einstein-Podolsky-Rosen channels", Phys. Rev. Lett. {\bf70} (1993) 1895.
\bibitem{AS2} H. Weng and S. Kais, "Quantum teleportation in one dimensional quantum dots system", Chem. Phys. Lett. {\bf421} (2006) 338.
\bibitem{AS4} T. M. Graham, H. J. Bernstein, T. C. Wei, M. Junge and P. G. Kwiat, "Superdense teleportation using hyperentangled photons", Nat. Commun. {\bf6} (2015) 7185.
\bibitem{AS5} R. Fortes and R. Gustavo, "Fighting noise with noise in realistic quantum teleportation." Phys. Rev. A {\bf92} (2015) 012338.
\bibitem{QW1} X. Tan, X. Zhang and J. Fang ``Perfect quantum teleportation by four-particle cluster state." Inf. Process. Lett. {\bf116} (2016) 347.
\bibitem{AS6} Y. Sun, X. Song, H. Qin, X. Zhang, Z. Yang and X. Zhang, "Non-local classical optical correlation and implementing analogy of quantum teleportation." Scientific Rep. {\bf5} (2015) 9175.
\bibitem{KL1} A. G. Adepoju, B. J. Falaye, G. H. Sun, O. Camacho-Nieto and S. H. Dong, ``JRSP of two-qubit equatorial state in quantum noisy channels", To appear in Phys. Lett. A (2017).
\bibitem{KL2} B. J. Falaye, G. H. Sun, O. Camacho-Nieto and S. H. Dong, ``JRSP of three-particle state via three tripartite GHZ class in quantum noisy channels", Int. J. Quantum Inform. {\bf14} (2016) 1650034.
\bibitem{AS7} H. O. Everitt, "Experimental Aspects of Quantum Computing", 2009, Springer, Us.
\bibitem{AS8} W. A. Coish and L. Daniel, "Quantum computing with spins in solids", Handbook of Magnetism and Advanced Magnetic Materials (2006)  (Preprint:	arXiv:cond-mat/0606550).
\bibitem{A2} X. Xu, Y. Wu, B. Sun, Q. Huang, J. Cheng, D. G. Steel, A. S. Bracker, D. Gammon, C. Emary and L. J. Sham, "Fast Spin State Initialization in a Singly Charged InAs-GaAs Quantum Dot by Optical Cooling", Phys. Rev. Lett. {\bf99} (2007) 097401.
\bibitem{A3} J. D. Mar, J. J. Baumberg, X. Xu, A. C. Irvine and D. A. Williams, "Ultrafast high-fidelity initialization of a quantum-dot spin qubit without magnetic fields", Phys. Rev. B {\bf90} (2014) 241303(R).
\bibitem{BJ2} M. A. Nielsen, I. L. Chuang {\it "Quantum computation and quantum information"}. Cambridge University Press, 2010.
\end{thebibliography}
\end{document}